\documentclass[runningheads]{llncs}

\usepackage[T1]{fontenc}
\usepackage{graphicx}
\usepackage{amsmath,amssymb}
\usepackage{booktabs}
\usepackage{multirow}
\usepackage{url}
\usepackage{listings}
\usepackage{xcolor} % for change-highlighting
\usepackage{algorithm}
\usepackage{algpseudocode}  % from algorithmicx

\usepackage{booktabs,tabularx,array,ragged2e,caption}
\usepackage{float}
\usepackage{stfloats}
\newcolumntype{Y}{>{\RaggedRight\arraybackslash}X}
\newcolumntype{C}[1]{>{\Centering\arraybackslash}p{#1}}

\usepackage[hidelinks]{hyperref}
\begin{document}

\title{Systematization of Knowledge: Formal Verification of Consensus Protocols}
\author{
Nikita Bondarev\inst{1,2}\thanks{Corresponding author.} \and 
Kirill Ziborov\inst{3,2} \and
Yury Yanovich\inst{1}
% \protect\footnotemark[2]
}

\authorrunning{N. Bondarev et al.}

\institute{
  Skolkovo Institute of Science and Technology, Moscow, Russia\\
  % \email{n.bondarev@skoltech.ru, y.yanovich@skoltech.ru}
  \and
  Lomonosov Moscow State University, Moscow, Russia\\
  % \email{krl.ziborov@gmail.com}
  \and
  Positive Technologies, Moscow, Russia
}

\maketitle

\begin{abstract}
\vspace{-0.5cm}
Formal verification is increasingly critical for blockchain consensus protocols, where subtle bugs can cause irreversible financial loss and network failure. Yet the literature on verification methods is fragmented across tools, protocol families, and property classes, hindering cumulative progress. This Systematization of Knowledge paper analyzes over 20 verified consensus protocols--from crash-fault-tolerant Raft to Byzantine-fault-tolerant HotStuff, DAG-based FairDAG, and proof-of-stake Beacon Chain--to establish a unified taxonomy of verification approaches. We introduce a verification maturity scale ranging from informal reasoning to machine-checked code proofs, and present a Protocol--Property--Method matrix mapping protocols to verified safety, liveness, and economic properties. Our analysis reveals persistent gaps: liveness verification remains underdeveloped despite its importance for progress guarantees; specification-implementation disconnects undermine real-world assurance; and scalability limits restrict verification to small networks. We provide practical recommendations for tool selection and proof engineering, and outline a research roadmap toward scalable, economically-aware verification. This work aims to guide both researchers and practitioners in building more rigorously verified consensus systems.

\end{abstract}

\vspace{-0.5cm}
\section{Introduction}
\label{sec:introduction}

Formal verification has become an indispensable discipline in the development of blockchain systems, where even minor implementation errors can lead to irreversible financial losses, network forks, and the erosion of user trust. As consensus protocols lie at the heart of every blockchain, ensuring their correctness under adversarial conditions is paramount. This Systematization of Knowledge (SoK) paper provides a comprehensive taxonomy of formal verification approaches applied to consensus protocols, analyzing both methodological foundations and practical outcomes across diverse protocol families. The necessity of rigorous verification is underscored by the high cost of failures in decentralized environments, where updates are difficult to revert and security guarantees must be mathematically sound rather than empirically tested.

Real-world incidents underscore the need for formal verification beyond testing: arithmetic overflows in the Beacon Chain~\cite{Cassez2022}, liveness violations in Tendermint Fastsync~\cite{Braithwaite2020}, and unfair reward distribution in federated learning consensus~\cite{Liu2025} all escaped conventional analysis. Yet existing surveys lack a unified taxonomy connecting protocol classes, verified properties, and verification methods~\cite{Verma2022,Bano2019,Raikwar2024}. This work addresses that gap by establishing a verification maturity scale and a Protocol--Property--Method matrix.

We analyze diverse consensus families: crash-fault-tolerant (Raft~\cite{Woos2016}), Byzantine-fault-tolerant (Tendermint~\cite{Maung2018}, HotStuff~\cite{Kukharenko2020}), proof-of-stake (Algorand~\cite{Alturki2020}, Beacon Chain~\cite{Afzaal2024}), and DAG-based protocols (FairDAG~\cite{DakaiKang2026}, AleoBFT~\cite{techrep-2025-aleobft}). Each presents distinct verification challenges addressed in our taxonomy.

This SoK makes several key contributions to consensus verification:
\begin{enumerate}
  \item A verification maturity scale from informal reasoning to machine-checked code proofs, enabling assessment of verification depth.
  \item A Protocol--Property--Method matrix mapping 20+ consensus protocols to verified safety, liveness, and fairness properties and their tools.
  \item Identification of critical gaps: liveness under asynchrony, dynamic stake handling, and the spec--implementation disconnect.
  \item Practical recommendations for tool selection and proof engineering, grounded in protocol characteristics and verification goals.
\end{enumerate}
These contributions systematize the state of the art to guide researchers and practitioners toward scalable, rigorous consensus verification.

\section{Formal Properties of Consensus}
\label{sec:properties}

Formal verification of consensus protocols is a mathematical proof that a system satisfies a certain set of properties even in the presence of failures or malicious participants. These properties are traditionally divided into safety, which ensures that nothing bad happens, and liveness, which ensures that something good eventually happens. However, modern blockchain protocols require additional guarantees beyond these classical properties, including fairness, economic incentives, and dynamic stake handling. In this section, we systematize the property classes that appear in the verification literature and explain their significance for different protocol families.

\subsection{Safety Properties}
\label{subsec:safety}

Safety properties form the foundation of consensus protocol verification. Formally, a safety property is an invariant that must hold at every step of the protocol's execution; its violation implies the occurrence of an invalid state. Under the Byzantine model, where nodes may behave arbitrarily, guaranteeing safety properties even in the presence of malicious participants is a minimal requirement for establishing trust in the system. From a practical perspective, safety properties are also more amenable to automated verification. Thus, formal verification of safety properties serves as a necessary first step toward building reliable consensus protocols. In this section, we systematize safety properties that are prioritized for verification in most papers.

At a higher level of abstraction, most safety properties verified in the literature can be unified as \textit{consistency safety}: honest participants must not produce incompatible protocol outcomes. In single-shot consensus, this appears as \textit{agreement} \cite{Rahli2018} on a single decided value, whereas in blockchain and state-machine-replication protocols it is typically formulated as prefix compatibility \cite{Praveen2024,JaskeliofF2025}, no-fork safety, or consistency of committed logs and finalized chains. For instance, in the Algorand protocol, the asynchronous safety property is proved, demonstrating that no two distinct blocks can be certified in the same round even with full adversary control over message delivery \cite{Alturki2020}.

Another safety-type property is \textit{validity}. It requires that the decided value originate from the protocol execution itself, rather than be fabricated or injected by a malicious participant. In the formal analysis of Tendermint, this appears as the impossibility of forcing honest nodes to accept an invalid block \cite{Maung2018}. In Red Belly Blockchain, a form of validity ensures that delivered values were broadcast by correct processes \cite{Tholoniat2019}.

For blockchain systems, classical safety properties are often extended with specific guarantees related to data structures and finalization. No-forking guarantees that the blockchain maintains a single canonical chain where conflicting branches cannot simultaneously reach finalization. In AleoBFT, the blockchain non-forking property is proved accounting for dynamic stake changes where validators can join and leave the network every block \cite{techrep-2025-aleobft}. For Raft replicated state-machine protocol safety is stated in a stronger, implementation-facing form: if one correct replica has applied a command at a log index, no other correct replica may apply a different command at that same index. Woos et al. \cite{Woos2016} lift that execution-level consistency to linearizability, meaning that client operations appear to take effect atomically in a single global order consistent with real-time completion. For Giskard, it is proved that blocks at each stage of the protocol cannot conflict, ensuring injectivity \cite{Li2020Giskard}. 

% PoS continue
In Proof-of-Stake protocols, \textit{justification} and \textit{finalization} are central safety notions used to express block irreversibility at the checkpoint level. A checkpoint is justified once it is endorsed by a supermajority of validators, while finalization additionally requires a justified descendant checkpoint, thereby turning local supermajority support into a stronger guarantee that the checkpoint belongs to the canonical history. In the Beacon Chain, these mechanisms are formally specified as the core of epoch-level finality \cite{Afzaal2024}. An even stronger notion, used in Gasper \cite{Casper}, is \textit{accountable safety}: if two conflicting checkpoints are ever finalized, then at least a threshold fraction of the total stake must have violated the protocol’s slashing conditions \cite{RV-Casper}. Thus, unlike ordinary safety, accountable safety not only rules out conflicting finalizations under the standard fault bound, but also ensures that any such violation leaves cryptographic evidence of slashable misbehavior.

\subsection{Liveness Properties}

Unlike safety properties, which guarantee the absence of invalid states, liveness properties capture system progress,  i.e., the requirement that "something good eventually happens". In the context of consensus protocols, this typically implies a guarantee that correct nodes will eventually reach a decision, known as \textit{termination}, that the protocol continues extending the ledger despite failures and adverse scheduling, referred to as \textit{progress}. In the surveyed literature, liveness verification is substantially less common than safety verification. The main reason is that liveness depends not only on the protocol logic itself, but also on temporal assumptions about timeouts, message delivery, scheduler fairness, and, in some cases, randomization. As a result, liveness proofs are typically more sensitive to the underlying execution model and are significantly harder to mechanize.

The liveness properties verified in the literature take several different forms depending on the protocol model and the proof objective. A standard formulation is eventual-progress liveness, which states that once the network becomes sufficiently well behaved, the protocol eventually makes progress and produces new decisions. This notion appears most clearly in the deductive verification of the Stellar Consensus Protocol by Losa and Dodds, who prove safety and liveness under eventual synchrony \cite{Losa2020}. A complementary result was obtained earlier by Yoo et al., who used model checking with UPPAAL to check both safety and liveness of SCP in a timed automata model, introducing explicit timer abstractions to keep timeout-sensitive verification tractable \cite{Yoo2019}. Another notion appearing in the literature is \textit{plausible liveness}, verified for Gasper Finality Gadget. This is more specialized property: rather than requiring that all fair executions eventually finalize, it states that from any reachable state there still exists a continuation in which checkpoints can again be finalized without violating slashing conditions, provided that more than two thirds of the stake follows the protocol. Accordingly, plausible liveness is better understood as a recoverability property for the finality gadget than as a full liveness theorem for the entire consensus stack.

Liveness proofs often require synchrony assumptions because deterministic consensus is impossible in a fully asynchronous network according to the classical Fisher-Lynch-Peterson result \cite{FLP}. Therefore, proofs often implemented on partial synchrony or specific adversary models. Such as the round-rigid adversary assumption used in verifying Red Belly Blockchain to guarantee fair scheduling of transitions and termination \cite{Tholoniat2019}. Randomization introduces an additional difficulty, because asynchronous protocols like HoneyBadgerBFT \cite{HoneyBadger} rely on random coins or probabilistic scheduling arguments to guarantee progress, but such mechanisms are difficult to encode in standard model checkers and are often unsupported directly. Additionally, the modelling complexity of modern protocols complicates liveness proofs. Tree-based models used in protocols like HotStuff make proving liveness more difficult, as ensuring tree structure and reachability predicates in inductive invariants increases verification complexity. As a result, many works focus solely on safety \cite{Kukharenko2020,Jehl2021}. 

% A strong motivation for formal liveness verification is the discovery of errors missed by manual proofs, because new counterexamples for known protocols having been identified through automated analysis. The possibility of infinite execution without termination was found for Casper FFG [Tholoniat2022]. Validators endlessly attempt to agree on a block at one level but cannot gather a quorum to move to the next. In HoneyBadgerBFT a liveness issue was identified in the randomized binary consensus. Adversary could prevent progress by controlling message delivery relative to the common coin algorithm's return value. Furthermore, model checking of the Tendermint Fastsync synchronization protocol using TLC revealed a termination violation counterexample, caused by the interdependence of timeouts and message delays \cite{Braithwaite2020}. This confirms that even industrial-grade protocols contain hidden liveness bugs that formal methods can uncover.

% A general approach to liveness verification remains an open problem in the field. While parametric and symbolic tools allow proving termination for certain classes, many modern protocols are verified only for safety. The discovery of counterexamples in protocols like Casper and HoneyBadger underscores the critical importance of automated liveness verification, as manual proofs often prove incomplete or erroneous under asynchrony and Byzantine faults.

\subsection{Protocol-Specific Properties}
\label{subsec:specific_properties}

Modern consensus protocols often require formal verification of properties that go beyond classical safety and liveness. These properties arise from blockchain-specific concerns such as transaction ordering, economic attacks, broadcast semantics, auditability, and the internal structure of protocol state spaces. Literature analysis reveals several key classes of such characteristics. They become objects of formal verification to ensure the reliability of blockchain systems.

One important class is \textit{transaction-ordering fairness}. In FairDAG, Kang et al.~\cite{DakaiKang2026} formalize Ordering Linearizability, which requires that if all correct nodes receive transaction $t_1$ before $t_2$, then $t_1$ will be ordered before $t_2$. They also verify $\gamma$\textit{-batch-order-fairness} for batch ordering. This property guarantees that if a fraction $\gamma$ of correct nodes receives $t_1$ before $t_2$, then $t_1$ will not be ordered later than $t_2$. These properties are specific to multi-proposer blockchain settings and are intended to limit manipulation of transaction ordering. A related but distinct property appears in Fino \cite{Malkhi2022}, which introduces  \textit{blind order fairness}: validators must not learn transaction contents before those transactions are irrevocably committed to a total order. Unlike ordinary fairness, this property targets resistance to MEV-style attacks, especially front-running based on transaction visibility at the consensus layer.

A second class concerns broadcast-level correctness properties. In the Red Belly protocol described in \cite{Tholoniat2019}, binary value broadcast properties are verified. These include the obligation property guaranteeing delivery if a sufficient number of nodes broadcast a value and uniformity guaranteeing that all correct nodes will deliver the value. These are protocol-specific guarantees for the correctness of the broadcast primitive on which consensus is built, rather than properties of the final consensus outcome itself. Closely related property is \textit{delivery integrity}, emphasized in recent work on DAG-based consensus \cite{Bertrand2024}. This property requires that a correct process neither delivers the same message twice nor delivers a message that was never correctly broadcast and rules out duplicates and phantom delivery.

Another protocol-specific direction concerns attack resilience and adversarial influence. In the formal analysis of Tendermint by Maung et al.~\cite{Maung2018}, the authors identify a censorship-related vulnerability: the protocol may fail to include valid transactions when a sufficiently large fraction of validators strategically refuses to vote. In a related direction, Esposito et al. \cite{Esposito2024} analyze coalition attacks against Algorand, showing that coordinated adversarial behavior can bias the protocol toward committing empty blocks once the adversarial coalition exceeds the admissible threshold.

Some verified properties are better viewed as auxiliary meta-properties rather than as safety, liveness, or attack-resistance guarantees. In the Coq formalization of CBC Casper, Li et al. prove \textit{non-triviality} and strong non-triviality \cite{Li2020a}. These results do not state that the protocol is safe or live; rather, they show that the protocol state space is sufficiently rich to admit genuinely branching behaviors, so that safety is not obtained vacuously from a degenerate model with no meaningful alternatives. Strong non-triviality strengthens this idea by giving a constructive witness to the branching structure of future executions. In the Agda verification of Chained HotStuff / LibraBFT, Carr et al. prove \textit{external verifiability}, meaning that an observer who does not participate in the protocol can still check the correctness of a committed result \cite{Carr2022}.

Overall, the spectrum of formally verified properties in consensus protocols is evolving.  In addition to safety and liveness, recent work verifies fairness of transaction ordering, privacy-preserving ordering guarantees, correctness of broadcast primitives, resistance to censorship and coalition attacks, and external auditability of committed outcomes. This shift reflects a broader change in the goals of formal verification: modern blockchain protocols must be shown not only to decide consistently and eventually make progress, but also to satisfy economic, structural, and adversarially robust guarantees that are specific to their intended deployment setting.

\begin{table}[htbp]
\centering
\caption{Classification of Verified Consensus Protocols by Category, Properties, Tools \& Modeling Approach}
\label{tab:protocol_classification}
\scriptsize
\begin{tabularx}{\textwidth}{|p{1cm}|p{2.8cm}|p{2.8cm}|X|p{1cm}|}
\hline
\textbf{Class} & \textbf{Algorithm} & \textbf{Verified Properties} & \textbf{Tools \& Modeling Approach} & \textbf{Ref.}\\
\hline

\multirow{4}{*}{\textbf{CFT}} 
& Raft & Safety & Rocq (Verdi) & \cite{Woos2016}\\
\cline{2-5}
& Paxos variants & Safety, Liveness & SPIN, Isabelle/HOL · Heard-Of Model & \cite{Tsuchiya2008,CharronBostMerz2009} \\
\cline{2-5}
& Ceph & Safety & TLA+/TLC & \cite{Fernandes2021} \\
\cline{2-5}
\hline

\multirow{12}{*}{\textbf{BFT}} 
& Ben-Or & Safety, Liveness & ByMC · Threshold automata, Round-Rigid Adversaries & \cite{Bertrand2019} \\
\cline{2-5}
& \multirow{2}{*}{Tendermint} & Safety, Liveness, Attack Resistance & PAT · CSP\# Process algebra & \cite{Maung2018} \\
& & Safety, Liveness  & TLA+/TLC, Apalache · Explicit \& symbolic MC & \cite{Braithwaite2020}  \\
\cline{2-5}
& \multirow{2}{*}{HotStuff} & Safety & Ivy, TLA+/TLAPS · FO model& \cite{Jehl2021}\\
& & Safety & TLA+/TLC & \cite{Kukharenko2020} \\
\cline{2-5}
& Chained HotStuff/LibraBFT & Safety, External Verifiability & Agda & \cite{Carr2022} \\
\cline{2-5}
& TetraBFT & Safety & TLA+/Apalache · Inductive invariants & \cite{Yu2024} \\
\cline{2-5}
& PBFT & Safety & Rocq (Velisarios) & \cite{Rahli2018} \\
\cline{2-5}
& Giskard & Safety & Rocq · Stage-based invariants & \cite{Li2020Giskard} \\
\cline{2-5}
& Streamlet & Safety & Agda · Executable spec & \cite{JaskeliofF2025} \\
\cline{2-5}
& Pipelined Moonshot & Safety & Ivy & \cite{Praveen2024} \\
\cline{2-5}
& Red Belly & Safety, Liveness, BV-Broadcast & ByMC · Parametric threshold automata & \cite{Tholoniat2019} \\
\cline{2-5}
& ChonkyBFT & Safety, Liveness & Quint, Apalache · Symbolic BMC & \cite{frança2025chonkybftconsensusprotocolzksync} \\
\hline

\multirow{6}{*}{\textbf{PoS}} 
& \multirow{2}{*}{Casper} & Safety, Liveness (FFG) & Rocq · Finality gadget model & \cite{RV-Casper} \\
& & Safety, Non-triviality (CBC) & Rocq · Layered abstraction & \cite{Li2020a}  \\
\cline{2-5}
& \multirow{2}{*}{Beacon Chain} & Safety & PAT · CSP\#, Epoch model & \cite{Afzaal2024} \\
& & Safety  & Dafny · Code-level deductive verif. & \cite{Cassez2022} \\
\cline{2-5}
& \multirow{2}{*}{Algorand} & Safety & Rocq · Timing \& delay model & \cite{Alturki2020} \\
& & Noninterference, Attack Influence & CADP/LNT · Noninterference analysis & \cite{Esposito2024} \\
\hline

\multirow{7}{*}{\textbf{DAG}} 
& DAG-Rider & Safety, Delivery Integrity & TLA+/TLAPS · DAG construction abstraction & \cite{Bertrand2024} \\
\cline{2-5}
& Hashgraph & Safety, Delivery Integrity & TLA+/TLAPS · DAG construction abstraction & \cite{Bertrand2024} \\
\cline{2-5}
& Aleph & Safety, Delivery Integrity & TLA+/TLAPS · DAG construction abstraction & \cite{Bertrand2024} \\
\cline{2-5}
& BullShark & Safety, Delivery Integrity & TLA+/TLAPS · DAG construction abstraction & \cite{Bertrand2024} \\
\cline{2-5}
& AleoBFT & Safety, Dynamic Stake & ACL2 · Dynamic stake model & \cite{techrep-2025-aleobft} \\
\cline{2-5}

\hline

\multirow{3}{*}{\textbf{Other}} 
& \multirow{2}{*}{Stellar SCP} & Safety, Liveness & UPPAAL · Timed automata & \cite{Yoo2019} \\
& & Safety, Liveness & Ivy, Isabelle/HOL · FO model + interactive proof of model soundness & \cite{Losa2020} \\
\cline{2-5}
& STBC & Safety, Trust Model & PAT · CSP\# Crowdsourcing model & \cite{Afzaal2022a} \\
\hline

\end{tabularx}
\end{table}

\section{Verification Methods and Abstraction Levels}
\label{sec:taxonomy}

In this section, we analyze verification approaches for consensus protocols along two dimensions: the methodological foundation (deductive, model checking, hybrid) and the level of abstraction at which verification is performed. This distinction matters because the choice of method and abstraction jointly determines which protocol classes can be analyzed, which properties can be established, and the trade-off between automation, scalability, and proof strength~(Table~\ref{tab:protocol_classification}).

\subsection{Interactive Theorem Proving}
\label{subsec:itp}
Deductive verification has been the most rigorous approach to work in machine-checked reasoning about consensus protocols. It typically requires translating the original algorithm, or its program implementation, into a formal model expressible in the target proof environment. Proofs are then carried out in proof assistants or solver-backed verification systems such as Rocq (formerly known as Coq), Isabelle/HOL, Agda, TLAPS, Ivy, or SMT-based frameworks, rather than by explicit state-space exploration. In this literature, the dominant proof style is inductive: authors define protocol invariants, refinement relations, or logical proof obligations and then show that every protocol step preserves them. The main advantage of this approach is that the resulting guarantees are usually parametric in the number of nodes and are not tied to a finite exploration bound, although this comes at a substantial proof-engineering cost.

A representative early result is the verification of Raft in Rocq within the Verdi framework by Woos et al \cite{Woos2016}. The authors provide the first machine-checked proof of state machine safety for Raft and connect it to an end-to-end guarantee of linearizable state-machine replication for the extracted implementation. The proof proceeds by reasoning about the evolution of the global protocol state using a large collection of inductive invariants. However, the verified properties remain fundamentally safety-oriented, and the work does not provide a liveness proof. A closely related effort is Velisarios by Rahli et al. \cite{Rahli2018}, who also use Rocq to verify PBFT. In contrast to Verdi, Velisarios is explicitly designed to model Byzantine malicious behavior, which makes it more directly applicable to BFT consensus protocols. 

Another line of work formally verifies the Casper family in Rocq. Li et al. \cite{Li2020a} formalize Correct-by-Construction Casper \cite{Casper} and prove safety together with non-triviality, showing how deductive methods can clarify the mathematical assumptions underlying the protocol. A subsequent effort by Runtime Verification \cite{RV-Casper} proves accountable safety and plausible liveness for Casper FFG. However, this liveness result should be interpreted carefully: it concerns the finality gadget, not consensus in the broader sense of a complete block production and fork-choice protocol.

Deductive verification has also been applied to several other blockchain consensus protocols.  Li et al. \cite{Li2020Giskard} specified Giskard in Rocq and formally verified its safety. For Algorand, Alturki et al. \cite{Alturki2020} develop a Rocq model that explicitly captures timing, network delay, and adversarial control of message delivery, and prove safety. For the Stellar Consensus Protocol, Losa and Dodds \cite{Losa2020} use Ivy to verify safety and some liveness properties for arbitrary but fixed quorum configurations in first order logic together with Isabelle/HOL used to prove the soundness of the first-order Ivy model. 

For the HotStuff family, the deductive literature develops in two main directions. Jehl \cite{Jehl2021} verifies a simplified version of HotStuff using Ivy and the TLA Proof System (TLAPS). The contribution is explicitly a proof of safety, and one of its main observations is methodological: HotStuff’s tree-shaped structure makes verification more difficult than the more traditional view-instance model used in earlier consensus protocols. A different approach is taken by Carr et al. \cite{Carr2022}, who formalize the Chained HotStuff/LibraBFT core protocol in Agda. Their proof establishes the protocol’s safety property and an additional condition enabling external verifiability of committed results by non-participating observers. They explicitly leave liveness outside the proved model, and their formalization covers only a single epoch or configuration, leaving reconfiguration for future work. A more recent solver-aided example is Praveen et al. \cite{Praveen2024} on Pipelined Moonshot, a HotStuff-style BFT protocol. Using Ivy, they prove safety, showing that as long as fewer than one third of validators are Byzantine, the protocol does not admit forks. 

A similar proof-assistant style appears in Jaskelioff et al. \cite{JaskeliofF2025}, who present an Agda formalization of the Streamlet consensus protocol. Their contribution is a mechanized proof of consistency, with particular emphasis on making the formalization both readable and computable, so that it can serve not only as a proof artifact but also as a basis for testing implementations.

More recently, deductive verification has started to move beyond one-off proofs of individual protocols toward reusable proof frameworks. In Bertrand et al. \cite{Bertrand2024}, the authors develop TLA+ specifications with TLAPS proofs for a family of DAG-based consensus protocols, including DAG-Rider, Cordial Miners, Hashgraph, Eventual Synchronous BullShark, and a variation of Aleph. The main verified property is safety. The proof is organized around reusable components for DAG construction and ordering, which reduces repeated proof effort across related protocols. This suggests that deductive verification may scale not only by proving isolated protocols correct, but also by developing reusable libraries of proof components.

Overall, the deductive literature is dominated by proofs of safety-style properties. Verified liveness is substantially less common, with Stellar and Casper FFG being notable exceptions; even there, the liveness result either depends on explicit synchrony assumptions or applies to a narrower finalization component rather than to the full consensus stack. We are not aware of deductive-only works in this set whose primary verified contribution is fairness or censorship resistance.

The strengths of deductive verification are clear: it provides the strongest form of machine-checked assurance and avoids the bounded-state limitations of model checking. In addition, code extraction from proof assistants into executable languages can narrow the gap between a verified formalization and a runnable artifact, and in some cases enables comparison with implementation-level code.  At the same time, the approach has significant drawbacks. First, it is highly labor-intensive, often requiring many auxiliary invariants and substantial proof engineering. Second, it scales poorly to liveness, since liveness proofs must encode subtle assumptions about time, fairness, scheduling, and partial synchrony. Finally, such proofs are often fragile under protocol evolution: relatively small changes in the protocol logic may require extensive proof revision. 

\subsection{Code-Level Verification}
\label{subsec:code_level}
Not all deductive work in the corpus operates at the level of a clean protocol theorem. Cassez et al. \cite{Cassez2022} verify a large, critical fragment of the Ethereum 2.0 Beacon Chain reference implementation in Dafny. Their main result is the verified absence of runtime errors together with additional functional-correctness specifications strong enough to uncover bugs and justify verified fixes. This is still deductive verification, but the verified properties are closer to software correctness and implementation safety than to consensus-level theorems.

\subsection{Model Checking}
\label{subsec:model_checking}

Model checking is an automated verification technique that involves the systematic exploration of the reachable states of a formal model to determine whether properties hold. Protocols are typically modeled as transition systems or finite-state abstractions, while properties are expressed in temporal logics such as Linear Temporal Logic (LTL) or Computation Tree Logic (CTL). Compared with interactive theorem proving, model checking requires much less manual proof effort and is especially effective at finding concrete counterexamples. This makes it particularly valuable in consensus verification, where subtle bugs often arise from rare interleavings of messages, faults, and timeouts. A representative example is the verification of Tendermint Fastsync, where TLC exposed a counterexample violating termination and helped identify a protocol flaw during refinement \cite{Braithwaite2020}.

The primary limitation of model checking is the exponential growth in the number of reachable states with an increase of number of processes, rounds, messages, and timer values, which usually restricts explicit-state verification to small configurations: often with $\leq$10 processes and without explicit modelling of timing \cite{Tsuchiya2008,Yoo2019,Kukharenko2020}. 

The analyzed works utilize a wide spectrum of model checking tools. Despite sharing a common goal---state space exploration---they differ in methods of state representation, supported logics, and scope of application. The choice of tool is dictated by the specifics of the protocol and scalability requirements.

Explicit State Checkers encompass tools that verify system correctness by explicitly storing and traversing every reachable state in memory to construct a reachability graph. While intuitive and effective for finding bugs in early design stages, these tools are inherently constrained by the state explosion problem. The primary representatives of this class include TLC (the model checker for TLA+), PAT (Process Analysis Toolkit), and SPIN (using the PROMELA language).

TLC is widely used for debugging distributed protocols due to its ability to generate concrete counterexamples. A notable application is found in the work of Braithwaite et al. \cite{Braithwaite2020} on the Tendermint Fastsync protocol. The authors developed a TLA+ multi-level specification and successfully employed TLC to identify a critical counterexample violating the termination property during the refinement phase. This demonstrated TLC's strength in catching logical errors early. However, the study also highlighted severe scalability limits: as the number of peers or the blockchain height increased, TLC rapidly exhausted computational resources. For complex liveness checks, the tool timed out after 24 hours, failing to complete an exhaustive search. 

The limitations of explicit checking were further underscored by Yu et al. \cite{Yu2024} in their verification of TetraBFT. Their initial attempt to verify safety using TLC failed even for a minimal configuration of 4 nodes (1 Byzantine), 3 proposed values, and 5 views, as the state space exploded immediately. 
A complementary case study is provided by Fernandes \cite{Fernandes2021} in his formal verification of the Ceph consensus algorithm, a Paxos-variant for distributed storage. The authors used TLC to verify safety properties (agreement and validity) for configurations up to 4 monitors and 2 proposed values. They successfully reproduced a known bug from an early Ceph version via counterexample generation, demonstrating TLC's practical utility for regression testing. To mitigate state explosion, the authors employed symmetry reduction and bounded constants. Despite these optimizations, verification for 4 monitors required processing 206 million states and took approximately 11 minutes, with exponential growth observed beyond this point. 

Similar scalability constraints appear in PAT-based verifications. PAT utilizes the CSP\# language, which uniquely combines high-level parallelism operators with low-level programming constructs like variables and arrays. Afzaal et al. \cite{Afzaal2022a,Afzaal2024} leveraged PAT to formally specify and verify consensus mechanisms for crowdsourcing (STBC) and the justification/finalization logic of the Ethereum Beacon Chain. The authors successfully proved absence of forks and fault tolerance guarantees. Despite these successes, empirical results revealed the classic bottleneck of explicit model checking: verification time and memory consumption exhibited exponential growth as the number of nodes or epochs increased. 

Comparable limits were observed with SPIN. SPIN, based on the PROMELA language, has been a staple for verifying consensus algorithms within the Heard-Of model. Tsuchiya et al. \cite{Tsuchiya2008} applied SPIN to verify Paxos-type algorithms (LastVoting and Hybrid-1), employing an optimization that abstracted away specific Heard-Of sets by representing all possible behaviors non-deterministically. Despite such optimizations, SPIN's scalability remains fundamentally limited by the number of processes. As documented in technical reports, even with aggressive state-space reductions, SPIN could only verify agreement for up to 3 processes. In contrast, approaches utilizing Bounded Model Checking (BMC) with SMT solvers extended this limit to 8--11 processes. 

Together, these case studies illustrate a consistent pattern across explicit checkers. The collective experience with TLC, PAT, and SPIN demonstrates that while explicit state checkers are powerful tools for debugging specifications and finding shallow bugs via counterexamples, they are ill-suited for proving correctness in realistic, large-scale blockchain environments. The exponential state growth restricts verification to toy configurations (often N $\leq$ 4), as seen in the TetraBFT, Tendermint, and Ceph case studies. Even with engineering optimizations---symmetry reduction, bounded constants, or custom visualizers---explicit enumeration cannot overcome the fundamental combinatorial barrier.

This limitation has driven a methodological shift in contemporary research: explicit checkers are increasingly used as a first step in a hybrid verification workflow, complemented by symbolic model checking or inductive invariant proofs to achieve scalable, parameterized guarantees.

Symbolic and SMT-based Checkers encompass tools that represent sets of states symbolically via logical formulas, delegating state exploration to SMT solvers (e.g., Z3, cvc5). This approach covers significantly larger state spaces than explicit enumeration. Prominent tools include Apalache \cite{Apalache} for TLA+ and ByMC \cite{ByMC} for parameterized threshold-guarded algorithms.

Bounded Model Checking (BMC) with SMT solvers has been pivotal for verifying consensus algorithms with large or infinite state spaces. Tsuchiya et al. \cite{Tsuchiya2008} reduced asynchronous consensus verification to single-phase problems solvable by SMT, scaling to $\sim$10 processes versus 3--4 for explicit checkers. Braithwaite et al. \cite{Braithwaite2020} showed Apalache finding Tendermint Fastsync errors in 10 minutes where TLC failed to terminate.

For parameterized correctness, ByMC leverages cutoff techniques and SMT solving. Tholoniat and Gramoli verified Red Belly Blockchain safety and liveness for arbitrary $N \geq 3F + 1$ \cite{Tholoniat2019}. 

Symbolic checkers handle larger parameter spaces with high automation, but face key constraints. BMC requires bounds on nodes/rounds. SMT solvers lack native support for probabilistic liveness, leaving randomized protocol termination unverified. Specifications outside decidable first-order fragments require manual abstraction.
Symbolic and SMT-based checkers significantly advance over explicit methods, enabling verification of deeper executions and larger configurations. However, bounded checks do not constitute full proofs, and probabilistic liveness remains out of reach. Modern research thus more often combines symbolic BMC with inductive invariant proofs or parameterized verification for comprehensive guarantees.

Beyond general-purpose checkers, specialized tools address domain-specific challenges. UPPAAL models timed automata to verify timeout-critical protocols like Stellar SCP, explicitly capturing asynchronous networks and synchronization for safety and liveness analysis~\cite{Yoo2019}. ByMC enables parameterized verification via threshold automata, abstracting pseudo-code into message-count guards to prove properties for arbitrary $n \geq 3f+1$ in round-based Byzantine protocols. While these tools trade generality for depth---enabling rigorous analysis of timing and scalability that generic checkers cannot efficiently handle---their applicability remains constrained by specific protocol structures.

\subsection{Hybrid Approaches and Emerging Frameworks}
\label{subsec:hybrid}

Hybrid verification has emerged as a natural direction for the formal analysis of consensus protocols, combining the automation of solver-based reasoning with the expressiveness of interactive proof. Such an approach makes it possible to specify protocols in a transition-system style amenable to automation, while retaining the ability to discharge difficult proof obligations manually when automation becomes insufficient. A recent example is Veil by Pîrlea et al. \cite{Pirlea2025}, a framework embedded in Lean for the automated and interactive verification of transition systems. Veil allows the user to describe a transition system and its specification in a simple imperative language, supports bounded model checking through Lean tactics, and generates verification conditions in first-order logic that are discharged automatically by SMT solvers. Its evaluation includes consensus-related case studies such as Paxos and Stellar, the latter verified within a single framework rather than through the cross-tool combination of Ivy and Isabelle/HOL used by Losa and Dodds \cite{Losa2020}. This illustrates the appeal of hybrid methods for consensus verification: they can reduce the trusted gap between modeling and proof while preserving a significant degree of automation. At the same time, as of March 2026, Veil framework does not yet support liveness verification natively.

\section{Abstraction Techniques for Scalability}
\label{sec:abstractions}

In this section, we discuss abstraction techniques used in the formal verification of consensus protocols. In practice, verification rarely targets the entire production implementation; instead, it is typically carried out over an abstract model of the protocol’s core logic or over a simplified protocol variant. This is particularly visible in the HotStuff literature: deductive proofs reason about an abstract single-epoch core independent of implementation~\cite{Jehl2021,Carr2022}, while model-checking work targets simplified TLA\texttt{+} specifications rather than production code~\cite{Kukharenko2020}. More broadly, most verification efforts omit cryptographic primitives or model them axiomatically, and treat network semantics separately from the protocol transition system itself. As a result, regardless of the verification methodology, one needs a sound operational abstraction that isolates the protocol logic from implementation details while preserving the properties of interest.

Approach-specific abstractions then build on top of this common operational layer. Deductive verification generally tolerates a lower level of abstraction than model checking, thanks to the richer semantics of proof assistants such as Rocq, Isabelle/HOL, and Agda. A particularly important technique is property-oriented abstraction hierarchy, where different properties are proved at different abstraction levels. The clearest example is CBC Casper, whose Coq development explicitly introduces several abstraction layers and uses them to separate the assumptions needed for safety from those needed for non-triviality \cite{Li2020a}. 

By contrast, the primary limitation of classical model checking is the exponential growth of the state space as the number of processes, message histories, or timer values increases. Time is a particularly acute source of blow-up. In consensus protocols, explicit modeling of timeouts can easily generate many states that differ only in clock valuations while contributing little to safety reasoning. For this reason, safety analyses often abstract timeouts into nondeterministic events, whereas liveness arguments usually require a more faithful treatment of the relationship between timeout expiration, message delay, and processing time. This tension is visible in work on Tendermint and SCP. Tendermint Fastsync specifications~\cite{Braithwaite2020} make clear that the consensus relies critically on timeout progression for liveness, which makes timer abstraction delicate; similarly, verification of SCP showed that timer- and delay-aware modeling is necessary in order to expose liveness-sensitive behaviors that disappear in purely untimed abstractions. Yoo et al. \cite{Yoo2019} therefore introduced explicit timing abstractions by parameterizing the maximum timer value and the gap between node clocks, allowing UPPAAL to check safety and liveness properties of SCP in a finite time while preserving timeout-dependent logic under different quorum configurations.

A different response to the scalability limits of model checking is to move from concrete process interleavings to parameterized symbolic models. The most influential example in the consensus literature is the use of threshold automata and the Byzantine Model Checker (ByMC) \cite{ByMC}. \textbf{Threshold automata (TA)} provide an abstraction for fault-tolerant distributed algorithms in which the global state is represented by counters over local control locations and transitions are guarded by threshold conditions over shared counters or parameters. This representation is especially well suited to consensus protocols whose control flow is driven by quorum predicates such as receiving at least $N-F$ messages, since these guards can be encoded directly without enumerating individual process identities. As a result, ByMC supports automatic parameterized verification for arbitrary number of processes and admissible fault thresholds, rather than for a few fixed instances. This abstraction has been used both for asynchronous randomized consensus, including Ben-Or’s and Bracha’s algorithms under round-rigid adversaries  \cite{Bertrand2019}, and for blockchain-style Byzantine consensus, most notably in the verification of Red Belly, where the protocol is decomposed into threshold-automaton models of inner broadcast and outer decision logic \cite{Tholoniat2019}. The main limitation of the approach is that its scalability relies on strong structural regularity: TA are most natural for symmetric, threshold-driven, round-based protocols, and become less natural for consensus designs with dynamic committees or highly asymmetric control flow. In addition, applying the technique in practice typically requires non-trivial manual modeling effort and substantial expertise in constructing sound and faithful threshold-automaton abstractions of the original protocol.

\subsection{Communication Model Abstractions}
\label{subsec:comm_abstractions}

Beyond threshold abstractions, another major family of communication-model abstractions is the \textbf{Heard-Of (HO) model} \cite{Charron-Bost2007}. In HO, the network is abstracted by the sets $HO(p,r)$ of processes from which process $p$ actually receives messages in round $r$. Timing assumptions, omission faults, and benign failures are then encoded not by explicit schedules of sends and receives, but by predicates over collections of heard-of sets. This shifts verification from low-level reasoning about message orderings to higher-level reasoning about round-based communication patterns. That abstraction proved effective for scalability: early work by Tsuchiya and Schiper \cite{Tsuchiya2008,Tsuchiya2011} used HO to reduce consensus verification to finite or phase-bounded obligations amenable to model checking, while Charron-Bost and Merz used Isabelle/HOL to verify Paxos at the HO level with a comparatively compact proof structure \cite{CharronBostMerz2009}. More recent work pursued cutoff-style reasoning, aiming to lift small-instance verification results to unbounded system sizes \cite{MaricSprengerBasin2017}, while HOME adapted the HO perspective to threshold-guarded BFT protocols through a dedicated modeling and verification environment \cite{Zhai2023}. At the same time, HO remains a specialized abstraction. It is most natural for fixed-process, round-based algorithms and considerably less natural for modern blockchain protocols with dynamic committees, threshold certificates, and DAG-based ordering. Thus, HO is a powerful but domain-specific abstraction: it yields scalability gains for well-structured consensus families but is less suited to contemporary blockchain protocols with dynamic committees or DAG-based ordering.

Another clear example of communication-level abstraction appears in \cite{Bertrand2024}, which distinguishes between reliable broadcast and unreliable communication specifications of DAG-based protocols. Authors do not model concrete network implementations; instead, they introduce abstract communication specifications chosen to preserve exactly the safety-relevant behavior needed for their proofs. For protocols based on reliable broadcast, they deliberately weaken the abstraction and retain only the integrity aspect of reliable broadcast, explicitly dropping validity and agreement because these are only needed for liveness and lie outside the scope of their safety proofs. For protocols based on unreliable communication, they abstract away implementation details even further: receiving a vertex is modeled as a process non-deterministically choosing a vertex from another process’s local DAG and copying it, thereby capturing all asynchronous interleavings that a concrete gossip- or broadcast-style implementation could produce. 

\subsection{State Management Abstractions}
\label{subsec:state_abstractions}

A distinct abstraction pattern in recent deductive verification is state management abstraction, where dynamic participation is modeled without introducing dynamic process creation or destruction into the formal semantics. A clear example is the ACL2 verification of AleoBFT by Coglio and McCarthy~\cite{techrep-2025-aleobft}, which targets a DAG-based BFT protocol with dynamic stake. Instead of modeling validators as entities that appear and disappear over time, the formal model keeps a static universe of validator addresses in the global state and derives the currently relevant validator sets from blockchain state. Committees are not stored as primitive state components; they are derived from the current blockchain prefix, with validator bonding and unbonding affecting committee membership only after a fixed lookback delay. This avoids explicit reasoning about process initialization and membership churn, while still capturing highly dynamic participation. Stake is handled similarly: bonding and unbonding are recorded in blocks, and the active committee with its stake distribution is computed from the resulting blockchain state. The proof obligation shifts from reasoning about reconfiguration events to reasoning about invariants over the mapping from blockchain state to validator sets. This makes the model well suited for inductive safety proofs, while preserving the essential feature that committees may change completely from block to block, subject only to the protocol's fixed lookback rule and the assumption that each derived committee contains less than one-third faulty stake.

\section{Research Challenges and Open Problems}
Our analysis identifies four persistent gaps that limit the impact of formal verification on blockchain consensus.

\textbf{Liveness and Economic Guarantees.} Safety verification dominates the literature, with only $\approx30\%$ of surveyed works addressing liveness. This imbalance stems from the inherent difficulty of modeling partial synchrony, timeout progression, and probabilistic scheduling---all required to bypass the FLP impossibility result. Beyond classical termination, modern protocols demand verification of transaction finality latency, censorship resistance, and MEV-resistant ordering fairness. These properties blur the line between functional correctness and economic performance, requiring frameworks that integrate distributed systems reasoning with game-theoretic models of rational adversaries. No existing tool handles dynamic stake, liveness, and economic incentives in a unified manner.

\textbf{Scalability and Parameterization.} State-space complexity severely restricts verification depth. Explicit checkers typically exhaust resources beyond 4--10 nodes, while symbolic BMC scales only marginally further. Parameterized verification via threshold automata elegantly handles symmetric, round-based protocols, but struggles with modern DAG-based designs and dynamic validator sets. Protocols like AleoBFT introduce cyclic dependencies between consensus logic and stake evolution, necessitating novel abstractions (e.g., lookback windows) to decouple membership changes from round boundaries. Automating these abstractions remains an open challenge.

\textbf{Specification--Implementation Gap.} Most verified models target abstract TLA+ or Rocq specifications that diverge significantly from production Go/Rust code. This disconnect creates assurance blind spots: verified theorems do not guarantee implementation correctness, and protocol upgrades quickly invalidate verification artifacts. While isolated efforts like Dafny-based Beacon Chain verification narrow this gap, systematic approaches---model extraction from code, verified code generation, and runtime monitoring---remain underdeveloped and difficult to integrate into continuous deployment pipelines.

\textbf{Methodological Fragmentation.} Verification efforts are predominantly one-off, using bespoke models and ad-hoc tool choices. Lack of standardized benchmarks, reusable component libraries (e.g., for reliable broadcast or view synchronization), and systematic abstraction guidelines hinders cumulative progress. The field needs shared infrastructure that enables empirical comparison of tools, lowers the expertise barrier, and supports incremental assurance from early design to production deployment.

\section{Recommendations and a Path Forward}
We propose a three-tier roadmap to advance consensus verification from isolated proofs to integrated engineering practice.

\textbf{Immediate: Pragmatic Tool Matching.} Practitioners should align verification depth with risk tolerance. Explicit model checking (TLC, SPIN) suits early-stage design exploration for $n \leq 4$; symbolic BMC (Apalache, Quint) scales to $\approx 10$ nodes for bounded safety; timed automata (UPPAAL) are essential for timeout-critical protocols. For unbounded guarantees or dynamic stake, deductive provers (Rocq, Agda, Isabelle) or hybrid frameworks (Veil, Ivy) are necessary despite higher proof-engineering costs. Code-level verification (Dafny) should target critical state-transition logic to bridge the spec--implementation gap.

\begin{table}[h]
\centering
\caption{Tool Selection Guide: Matching Protocol Characteristics to Verification Methods}
\label{tab:tool_selection}
\scriptsize
\begin{tabularx}{\textwidth}{p{3cm}|X|p{1.7cm}|p{4cm}}
\toprule
\textbf{Protocol Feature} & \textbf{Recommended Approach} & \textbf{Tools} & \textbf{Rationale / Limitations} \\
\midrule

Early design analysis, bug-finding & Explicit-state Model Checking & TLC, PAT, SPIN & Fast feedback on small configs ($n \leq 4$); not exhaustive for large systems \\
\midrule

Deep simulations, inductive invariants & Symbolic MC & Apalache, Quint & Better scaling than explicit (up to $n \approx 10$); requires TLA+ or SMT encoding \\
\midrule

Timeouts, real-time logic & Timed Automata MC & UPPAAL & Explicit time modeling; state explosion for large configs \\
\midrule
Arbitrary $n$, threshold guards & Parametric MC (Threshold Automata) & ByMC & Proves for any $n \geq 3f+1$; requires symmetric, round-based structure \\
\midrule

Parametric safety, Complex invariants: dynamic stake, economic properties,
adversarial strategies 
& Deductive verification in ITP or hybrid frameworks & Rocq, Isabelle/HOL, Agda, Ivy, Veil & Strongest guarantees; high proof-engineering cost, liveness hard \\
\midrule

Implementation-level correctness & Deductive Code Verification & Dafny & Closes spec-impl gap; labor-intensive, protocol-agnostic properties \\
\bottomrule
\end{tabularx}
\end{table}

To ensure maintainability, adopt modular proof engineering: separate communication assumptions from ordering logic, define interface lemmas for local reasoning, and invest in reusable specifications for primitives like reliable broadcast and threshold signatures. Where possible, refine abstractions incrementally---start with high-level models (e.g., Heard-Of) for scalability, then concretize for implementation verification.

\textbf{Medium-Term: Standardized Infrastructure.} Develop integrated pipelines combining specification-guided fuzzing, symbolic execution, and deductive proofs for incremental assurance. Prioritize benchmark suites spanning CFT, BFT, PoS, and DAG families with reference implementations, formal specs, and known vulnerabilities. Automated abstraction synthesis---using program analysis or ML to derive threshold or HO abstractions from code---would lower the expertise barrier significantly.

\textbf{Long-Term: Unified Economic-Security Frameworks.} Future work must merge safety, liveness, and economic rationality into single verification frameworks. This requires extending temporal logic with probabilistic and game-theoretic semantics to reason about strategic deviations, MEV resistance, and finality latency under rational adversaries. The ultimate goal is certified end-to-end implementations: domain-specific consensus languages compiled to verified bytecode, backed by proof-carrying code and formally verified runtime systems. Modular proof engineering and reusable component libraries will be essential to scale this vision.

\section{Conclusion}
This SoK systematizes formal verification approaches for blockchain consensus, mapping over thirty protocols to a unified taxonomy of properties, methods, and abstractions. While significant progress has been made in verifying safety for major protocols like Raft, Tendermint, and Algorand, critical gaps persist. Liveness verification remains underdeveloped, scalability constraints limit analysis to toy configurations, and verified models frequently diverge from production code. Moreover, the lack of standardized tooling and reusable abstractions hinders cumulative progress.

Addressing these challenges requires a paradigm shift: from isolated, safety-focused proofs to scalable, economically-aware verification integrated into the protocol development lifecycle. By adopting pragmatic tool-selection strategies, investing in modular proof engineering, and developing unified frameworks for dynamic stake and rational adversaries, the community can deliver the rigorous guarantees demanded by decentralized systems securing billions in value. Formal verification must evolve from a post-hoc audit tool to a foundational engineering practice.

\bibliographystyle{splncs04_custom}
\bibliography{refs2025}

\end{document}